\documentclass[12pt,preprint]{aastex}

\newcommand{\lta}{{\>\rlap{\raise2pt\hbox{$<$}}\lower3pt\hbox{$\sim$}\>}}
\newcommand{\gta}{{\>\rlap{\raise2pt\hbox{$>$}}\lower3pt\hbox{$\sim$}\>}}

\usepackage{graphicx} 
\usepackage{amssymb}

\begin{document}

\title{The Environments of High Redshift QSOs}
\author{Soyoung Kim\altaffilmark{2}}
\email{sykim@pha.jhu.edu}
\author{Massimo Stiavelli\altaffilmark{2,3}}
\email{mstiavel@stsci.edu}
\author{M. Trenti\altaffilmark{2}}
\author{C.M. Pavlovsky\altaffilmark{2}}
\author{S.G. Djorgovski\altaffilmark{4}}
\author{C. Scarlata\altaffilmark{5}}
\author{D. Stern\altaffilmark{6}}
\author{A. Mahabal\altaffilmark{3}}
\author{D. Thompson\altaffilmark{3}}
\author{M. Dickinson\altaffilmark{7}}
\author{N. Panagia\altaffilmark{8}}
\author{G. Meylan\altaffilmark{9}}

\altaffiltext{1}{Based on observations with the NASA/ESA {\it Hubble Space Telescope},
obtained at the Space Telescope Science Institute, which is operated by the Association of
Universities of Research in Astronomy, Inc., under NASA contract NAS5-26555}

\altaffiltext{2}{The Johns Hopkins University, 3400 N. Charles St.,Baltimore, MD 21218}
\altaffiltext{3}{Space Telescope Science Institute, Baltimore, MD 21218}
\altaffiltext{4}{California Institute of Technology,
                 MS 105-24, Pasadena, CA 91125}
\altaffiltext{5}{Spitzer Science Center, Pasadena, CA }
\altaffiltext{6}{Jet Propulsion Laboratory, California Institute of Technology,
                 Mail Stop 169-506, Pasadena, CA 91109}
\altaffiltext{7}{National Optical Astronomical Observatories,
                 P.O. Box 26732, Tucson, AZ 85726}
\altaffiltext{8}{Space Telescope Science Institute, Baltimore, MD 21218;  Supernova Ltd., Olde Yard Village 131, Northsound Road, Virgin Gorda, British Virgin Islands}
\altaffiltext{9}{Laboratoire d'Astrophysique, Ecole Polytechnique F\'ed\'erale de Lausanne (EPFL)
                 Observatoire de Sauverny, CH-1290 Versoix, Switzerland }

\begin{abstract}
We present a sample of $i_{775}$-dropout candidates identified  in five
Hubble Advanced Camera for Surveys fields centered on Sloan Digital
Sky Survey QSOs at redshift $z\sim 6$. Our fields are as deep as
the Great Observatory Origins Deep Survey (GOODS) ACS images which
are used as a reference field sample. We find them to be overdense in two fields,
underdense in two fields, and as dense as the average density of
GOODS in one field. The two excess fields show significantly
different color distributions from that of GOODS at the 99\%
confidence level, strengthening the idea that the excess objects are
indeed associated with the QSO. The distribution of $i_{775}$-dropout
counts in the five fields is broader than that derived from GOODS at
the 80\% to 96\% confidence level, depending on which selection criteria
were adopted to identify $i_{775}$-dropouts; its width cannot be explained by cosmic
variance alone. 
Thus, QSOs seem to affect their environments in complex ways. We suggest the picture
where the highest redshift QSOs are located in very massive overdensities and are
therefore surrounded by an overdensity of lower mass halos.
Radiative feedback by the QSO can in some cases prevent halos from
becoming galaxies, thereby generating in extreme cases an
underdensity of galaxies. The presence of both enhancement and
suppression is compatible with the expected differences between lines
of sight at the end of reionization as the presence of residual
diffuse neutral hydrogen would provide young galaxies with shielding
from the radiative effects of the QSO.
\end{abstract}


\section{INTRODUCTION}

Observational astronomy has finally reached the point of beginning
to probe the era of reionization of hydrogen. The long search for
Gunn-Peterson \citep{Gunn65} troughs in the spectra of increasingly
higher redshift QSOs has finally become fruitful with the Sloan
Digital Sky Survey (SDSS). A dramatic increase in the intergalactic
hydrogen absorption at $z\simeq6$ was detected in the spectra of
high-redshift SDSS QSOs \citep[e.g.,][]{Becker01,Djorgovski01, White03}.
This was followed by the possible detection of a Gunn-Peterson
trough in the spectrum of QSO SDSS J1030+0524 at $z=6.28$
\citep{Fan01}. The case for termination of the reionization epoch at
$z\sim6$ is now relatively solid \citep[e.g.,][]{Fan06}, even if
not universally agreed upon \citep[e.g.,][]{Lidz06, Bolton07}. At
the same time, the Compton optical depth $\tau=0.084\pm0.016$ from
the five year WMAP data \citep{Komatsu08} is compatible with a
somewhat extended reionization process terminating at $z\simeq6$
\citep[e.g.,][]{Shull07}.

Despite the growing consensus that reionization may have terminated
at $z\simeq6$, it is extremely  unlikely that it occurred in a
universally synchronized fashion. Fluctuations from line of sight to
line of sight are generally expected due to clumpiness of the IGM,
and the gradual development and clumpy distribution of the first
ionizing sources, either proto-galaxies or early AGN
\citep[e.g.,][]{Miralda00}. Thus, reionization is expected to occur
gradually as the UV emissivity increases \citep[cf.][]{McDonald01},
with the lowest density regions becoming fully reionized first. This
is also suggested by modern numerical simulations
\citep[e.g.,][]{Ciardi03,Gnedin97,Gnedin04} which predict an
extended period of reionization, starting at $z \sim 15$ or even
higher and ending at $z\sim 6$ \citep[see
also][]{Cen03,Haiman03,Somerville03,Wyithe03}.

If reionization is completed at $z\simeq6$, it is reasonable to
attempt to identify the galaxies responsible for it. The combined
Great Observatory Origins Deep Survey \citep[GOODS;][]{Giavalisco04}
and Hubble Ultra Deep Field \citep[HUDF;][]{Beckwith06} have provided a
large sample of $i_{775}$-dropout galaxies. Unfortunately, their estimated
ionizing flux is insufficient to reionize the universe under
standard assumptions \citep{Bunker04,Dickinson04,Bouwens07}; one
would have to assume top heavy, very metal-poor stellar populations
\citep{Stiavelli04}, or rely on a burgeoning population of dwarf
galaxies brought about by a steep faint end slope of the luminosity
function \citep{Yan04}. A last alternative is that reionization was
very gradual \citep[e.g.,][]{Bouwens07}. Unfortunately, testing
these ideas is observationally challenging. At the same time, given
the predominance of the HUDF data on the derivation of the faint end
luminosity function of $i_{775}$-dropouts, one is led to wonder how much
these results are affected by cosmic variance given the small volume
probed by the HUDF. On this issue, conflicting claims regarding the
density of HUDF $i_{775}$-dropouts can be found in the literature, with
\citet{Bouwens07} arguing in favor of an underdensity \citep[see
also][]{Oesch07} while \citet{Malhotra05} argued in favor of an
overdensity \citep[however, see][]{TS08}.

In general, one would expect very high redshift galaxies to be highly clustered, especially
if purely gravitational clustering effects were amplified by positive feedback.
Thus, in order to address the importance and sign of feedback in the environments where
they should be easiest to detect, we were led to focus on fields centered on $z\gta6$ QSOs
as they should be the most clustered environments at
these very high redshifts and the strongest cases of feedback available for study.

Indeed, a generic expectation in most models of galaxy formation is
that the most massive density peaks in the early universe are likely
to be strongly clustered \citep{Kaiser84, Efstathiou88}. The
evidence for such bias is already seen with large samples of
Lyman-break galaxies at $z\sim3 - 3.5$ \citep{Steidel03}, and in
Lyman $\alpha$ selected galaxy samples \citep[e.g.,][]{Venemans03, Ouchi05},
and it should be even stronger at higher redshifts. An excess in the
number of galaxies and in the density of star formation was also
discovered in a systematic Keck survey of fields centered on known
$z>4$ quasars \citep[e.g.,][]{Djorgo99, Djorgovski99,Djorgovski03}.
The high metallicity associated with QSOs \citep{Barth03} -- even at $z \gta 6$ --
is often interpreted as evidence that they are located at the center
of massive (proto--)galaxies, thereby corroborating the overall
picture.  These arguments justify the expectation that QSOs at $z
\simeq 6$ most likely highlight some of the first perturbations that
become non--linear in the density distribution of matter \citep[see
e.g.,][]{TS07}.

However, QSOs are not ``quiet neighbors''. The intense emission of
ionizing radiation associated with QSOs ionizes the surrounding IGM
and may even photo-evaporate gas in neighboring dark halos before
this has an opportunity to cool and form stars \citep{Shapiro01}. In
this context, QSOs would suppress galaxy formation in their
vicinities. One would then observe a paucity of galaxies near a QSO
despite the underlying excess of dark halos. Moreover, near the reionization epoch 
the fraction of neutral hydrogen in the IGM may change rapidly, possibly shifting the
balance of the two effects. It would be exciting to see a change
from source enhancement to suppression around reionization by
observing a sample of $z=6$ QSOs.

It is with this goal in mind that we started a study of the
environment of  the five then known QSOs at $z \gtrsim 6$ using the
Advanced Camera for Surveys (ACS) on board the Hubble
Space Telescope (HST) to obtain images in the F775W
($i_{775}$) and in the F850LP ($z_{850}$) filters so as to identify
candidate objects at $z=6$ as $i_{775}$-dropout galaxies. All five fields
were observed to the same depth as GOODS in the $i_{775}$ and
$z_{850}$ bands so that GOODS can be used as a reference field
sample.

In a previous paper \citep{Stiavelli05}, we analyzed the number of
$i_{775}$-dropout galaxies identified in a HST/ACS field centered on the SDSS QSO J1030+0524
at $z=6.28$. In this field we found a very significant excess of sources compared
to the density of $i_{775}$-dropouts seen in GOODS, thus suggesting that clustering
wins over negative feedback. \citet{Zheng06} also observed a radio-loud QSO at $z\sim6$, SDSS J0836+0054, using ACS and detected a significant overdensity of i-dropout galaxies in its vicinity.
In this paper, we analyze four additional QSO fields in order to test and expand this result.

Section 2 is a description of the observations and data analysis.
Section 3 describes our $i_{775}$-dropout objects and their properties.
Section 4 contains discussion of our results and Section 5 summarizes our conclusions.
In this paper we use AB magnitudes and assume the cosmological parameters,
 $H_{0}= 70\, {\rm km}\, {\rm s}^{-1}\, {\rm Mpc}^{-1}$, $\Omega_{m}= 0.26$, and $\Omega_{\Lambda}= 0.74$.

\section{DATA REDUCTION AND ANALYSIS}

We observed five fields centered on five SDSS QSOs at redshift $z\gta6$ 
with the ACS/WFC on board HST. The QSOs were the most distant quasars known at the
time of our original Cycle 12 proposal. All are radio-quiet. Our targets were SDSS
J1148+5251 at $z=6.40\pm0.01$ \citep{Barth03}, SDSS J1030+0524 at $z=6.28\pm0.03$, SDSS J1306+0356 at $z=5.99\pm 0.03$, SDSS J1048+4637 at $z=6.23\pm0.03$, and SDSS J1630+4012 at $z=6.05\pm0.03$ \citep{Fan01, Fan03}. 

Table 1 summarizes the observations. Our observations in the F775W
($i_{775}$) and the F850LP ($z_{850}$) filters were designed to have similar exposure times to those used for the original (version 1.0) GOODS data products. The data were processed by the ACS pipeline CALACS that carries out bias and dark current removal and flat-fielding. The
individual calibrated images ({\it flt} files) were combined into a
single image for each filter using Multidrizzle, a pyraf application
based upon the drizzle algorithm \citep{Fruchter02}. Drizzle also
requires weight maps which we computed following the same procedure as was used for the GOODS data reduction:
\begin{equation}
Variance = \frac{\left[ (Dt+fB) + \sigma_{read} ^{2}
\right]}{(ft)^{2}}
\end{equation}
\begin{equation}
Weight = \frac{1}{(Variance)}
\end{equation}
where $D$ is the dark current (electron/sec/pixel), $f$ is the pixel
value of the reference flat field, $B$ is the background (electron/pixel) measured
in flat-fielded images, $t$ is the exposure time (second), and
$\sigma_{read}$ is the read-out noise (electron/pixel). We ran
MultiDrizzle (Koekemoer et al. 2002) with parameters {\it pixfrac}=
1.0, {\it final\_scale}= 0.03 and {\it final\_wht\_type}= {\it ivm} (individual weight map). 
The area of the final images is approximately 11.3 arcmin$^{2}$.
We measured the actual background noise in the drizzled ACS images, measuring
and correcting for the correlation between pixels introduced by the drizzling and
resampling process, and compared this to the variance predicted by the noise model
used to generate the weight maps (equations 1 and 2).  This correction was also verified
by block averaging the images and measuring the resulting noise directly on scales
larger than the inter-pixel correlation lengths.  The variance maps were adjusted
using this correction, and converted to rms maps which were provided to SExtractor \citep{Bertin96}
to modulate the source detection thresholds and to compute photometric uncertainties.

The catalogs were obtained using SExtractor, run on
the drizzled science images and with the same input parameters as
those for the GOODS catalogs (for both the HDFN and the CDFS). We applied the same procedures to all five
fields. The $z_{850}$ band images were used as the detection images
when running SExtractor in dual-image mode. We required objects to be
detected at a signal-to-noise (S/N)$>5$ in the $z_{850}$ band.
For the total magnitude of a source, we adopted SExtractor's
MAG\_AUTO values. The adopted magnitude zero points were 25.6405 and
24.8432 in $i_{775}$ and $z_{850}$, respectively. We computed
$i_{775}-z_{850}$ colors using the MAG\_ISO values to compare the
same isophotes in the two bands. 
For $i_{775}$ band sources detected
at less than the two sigma level in isophotal apertures, we computed
lower limits for the colors using the 2$\sigma$ upper limit to the $i_{775}$ band isophotal flux. 
The Galactic extinction estimate of E(B-V) was obtained from \citet{Schlegel98}
for GOODS and each QSO field. We determined the corrections for the $i_{775}$ and $z_{850}$ magnitudes using SYNPHOT. The actual corrections in the
two bands were as follows: 0.024 and 0.018 for HDFN, 0.016 and 0.012 for CDFS,
0.048 and 0.036 for J1030+0524; 0.022 and
0.016 for J1630+4012; 0.036 and 0.027 for J1048+4637; 0.044 and
0.033 for J1148+5251; and 0.060 and 0.045 for J1306+0356. The
limiting magnitudes and completeness levels were comparable to those
of GOODS catalogs.

\section{CANDIDATE OBJECTS}

The selection criteria are based on the $i_{775}-z_{850}$ color,
a magnitude limit $z\leq26.5$,  limits on S/N ratios, and the
SExtractor extraction $flag=0$ which identifies non-saturated and isolated sources outside the masked zones. We have considered two different
values of S/N$=5$ and $8,$ and the color limits of $i_{775}-z_{850}=
1.3$ and $1.5$. Objects selected with S/N$>5$ and
$i_{775}-z_{850} > 1.3$ will constitute our least restrictive sample S1; objects with
S/N$>5$ and $i_{775}-z_{850} > 1.5$ are our sample S2; and those
with S/N$>8$ and $i_{775}-z_{850} > 1.3$ are our sample S3. We eliminate objects that reside near the edges and on the star
diffraction spikes, as well as objects that appear to be artifacts
during visual inspection. GOODS candidates were selected by the same
selection criteria using the GOODS catalogs (version 1.1), including visual inspection. 
However, as the QSO fields only have ACS imaging in two bands, we do not require non-detections ($<2 \sigma$) in the $B_{435}$ and $V_{606}$ as was implemented in the \citet{Dickinson04} selection of  $i_{775}$-dropouts in the GOODS fields. Therefore, our GOODS $i_{775}$-dropout sample is different from the one used in \citet{Dickinson04}. Table~2 shows the number of $i_{775}$-dropouts selected in QSO fields and GOODS for different S/N ratios and color limits. 
In Table~2, the number of $i_{775}$-dropouts in GOODS is normalized
to the area of a single ACS/WFC field ($\sim$ 11.3 arcmin$^{2}$). 
The measurements of all quasar field candidates with $i_{775}-z_{850}
> 1.3$ and S/N$>5$ are listed in Table~3.

Contamination by stars is a potential concern. We estimated a priori
the possible contamination from stars by using as a proxy the number
density of stars brighter than visual magnitude $m_{v}=21$ at the
Galactic latitude of the five QSO fields. All fields have lower star
density than the mean star density at the galactic latitude of each QSO \citep{Zombeck90} at the galactic latitude of each QSO. In particular, the J1030+0524
field has a lower star density than GOODS, while the other
overdense field, J1630+4012, has a star density 4.8 times higher than GOODS. 
This suggests a degree of caution is necessary in excluding stars. We have
identified stars using the SExtractor star-galaxy index, S/G, half-light radius, $r_{hl}$ and $z_{850}$ mag. The criteria for stars were S/G$\geq0.85$, $r_{hl}\leq0.1$
arcsecond, and $z_{850}<25.5$ applied to the S1 samples. We found no stellar $i_{775}$-dropout candidates in our five fields but found 16 stellar $i_{775}$-dropout candidates (0.55 stars per ACS field) in GOODS.

Our target QSOs are not all flagged as stars because of the long wavelength point source halo effect seen with ACS. The point spread function in the F850LP filter is characterized by a long
wavelength halo which is due to light traveling through the CCD,
bouncing off the front side at a large angle, going once again
through the CCD and being detected. This effect is very
wavelength-dependent (and thus, for high-redshift QSOs, redshift-dependent).
Well-exposed images of a QSO will show this extended halo and the
QSO will fail to be identified as a star. The same would be true for
very red stars. However, if we artificially dim the QSOs to have similar apparent magnitudes as the other $i_{775}$-dropouts, the halos drop below the noise level and the fainter versions of our QSOs are identified as stars.

We also estimated the possible contamination by stars fainter than 25.5 by considering the
candidates with S/G$\geq0.85$, and half light radius $r_{hl}\leq0.1$
arcsecond. In Table 3, we have two objects (A8 and B2) in J1030+0524
and J1630+4012 that satisfy this relaxed criteria. When applied to GOODS, we found 11 (very red) objects (0.38 objects per ACS field) out of 235 objects selected using the S1 criteria.

For S/N$> 5$ and $i_{775}-z_{850}>1.3$ (our selection S1) we
see that two fields, J1030+0524 and J1630+4012, show an overdensity;
J1048+4637 has approximately the same number density of $i_{775}$-dropouts as GOODS; and the
J1148+5251 and J1306+0356 fields appear underdense compared to GOODS.

We have verified whether the variations in the number of candidates could be due to 
field-to-field background noise variations. We find  these variations to be generally 
small and that the background noise is highest in the field of J1030+0524, i.e., the one 
with the largest  excess. Thus, we conclude that background noise variations are not  
affecting our results.

Figures 1 through 5 show for each field the number counts as a
function of the $z_{850}$ magnitude (panel a) and as a function of
$i_{775}-z_{850}$ color (panel b). Panel c shows the count
distribution as a function of magnitude for objects redder
than $i_{775}-z_{850}=1.5$ and panel d shows the number of objects
redder than a given $i_{775}-z_{850}$ color in 0.1-magnitude bins for $i_{775}-z_{850}>0.9$.
The solid line shows the data for galaxies in the QSO fields.
The dotted line shows the distributions for the GOODS fields.

Figure~1-(d) shows the color distribution of galaxies in the 
J1030+0524 field (excluding the QSO) and GOODS. Their
distributions appear to be different, especially around $i_{775}-z_{850}\sim 2$.
In Figure~2-(d), the color distributions of J1630+4012 and GOODS appear
to be different for $i_{775}-z_{850}>1.7$. 
We applied the Chi-square ($\chi^2$) test on the binned color distributions to determine the significance of the differences between the color
distributions of the QSO fields compared to GOODS. We
focused on sources with S/N $>5$ that fall in the color interval
$1.3<i_{775}-z_{850}< 2.6$. For J1030+0524, the chi-square test yielded
a $\chi^2$ statistic of 30 and a probability of $P=0.3$\%
where P is the one-tailed probability that obtains a value of $\chi^2$ or greater --- e.g., there is less than a 0.3\% chance that both the GOODS and J1030+0524 $i_{775}$-dropout samples were drawn from the same distribution over the color range considered. 
For the other overdense field, J1630+4012, we found $\chi^2=52$ and $P<0.1\%$. 
For the other three fields, $\chi^2= 11$ and $P=41$\% for J1048+4637, $\chi^2=7$ and $P=83$\% for J1148+5251, and $\chi^2=7$ and $P=83$\% for J1306+0356.
For two overdense fields, the probability is not more than 0.3\% regardless of the specific criterion we use (S2 and S3 samples). 
Thus, our candidates in both overdense fields
have significantly different color distributions compared to GOODS.

Figure~6 shows substantial
spatial clustering of the $i_{775}$-dropout candidates in the J1030+0524 field:
 when the field is divided in half across the diagonal, almost all of the
sources are in the south-west half of the field. This makes the
excess in J1030+0524 even more significant. The color magnitude
diagram of candidates listed in Table 3 is presented in Figure~7, showing that
the overdense fields have fainter $i_{775}$-dropouts than GOODS. It is
notable that \citet{Willott05} in their less sensitive survey for $i_{775}$-dropouts around high-redshift SDSS QSOs, including J1030+0524, found no overdensities. The upper panel of Figure~8 shows
half-light radius versus $z_{850}$ for the $i_{775}$-dropout candidates from GOODS and the QSO fields. There is an upper envelope to the size-magnitude
relation, and the bottom panel of Figure~8 shows a histogram comparing the size distribution of GOODS and QSO field $i_{775}$-dropout half-light radii. It appears that the candidates in the overdense fields are more compact than those in GOODS, but this is not statistically significant.

\section{DISCUSSION}

Despite a complete reanalysis and a change in the type of SExtractor
magnitudes used to compute the $i_{775}-z_{850}$ color for dropout selection (from AUTO
to ISO mags), we confirm the overdensity in the J1030+0524 field
reported in \citet{Stiavelli05}. 
The overdensity is significant not only in the counts by
themselves but also in the color distribution.
Indeed, the departure of the color distribution of J1030+0524 and J1630+4012 is in the
sense of having an excess of red dropouts with precisely the colors
that one would expect from objects at the redshift of the two QSOs.
This makes the excess even more convincing.

One uncertain component of the comparison with GOODS is the
possible contamination by low redshift and Galactic interlopers.
Figure~9 shows the fraction of GOODS $i_{775}$-dropout objects selected by us but rejected when using the full GOODS $i_{775}$-dropout criteria including the $V_{606}$ data \citep{Beckwith06} to the number of GOODS $i_{775}$-dropouts selected by our criteria vs. the $i_{775}-z_{850}$ color. 
At $i_{775}-z_{850} > 1.7$, where the excess of $i_{775}$-dropouts is large in the J1030+0524 and J1630+4012 fields, there is less than 15\% contamination from potential foreground objects. Statistically, the full GOODS criteria would remove more objects from GOODS than from the J1030+0524 or J1630+4012 field because the latter have a redder color distribution. Thus, we do not think that the detected excess is due to interloper contamination.

In order to understand how unusual it is to identify this distribution
of over- and underdensities, we consider the number of $i_{775}$-dropouts
identified in 30 distinct and non-overlapping ACS fields in GOODS.
Figure~10 presents the resulting histogram of the number of
$i_{775}$-dropouts identified per unique GOODS ACS field using the
S1 and the S2 selection criteria. These distributions are reasonably
well fit by Poisson distributions with a mean of 6.5 (3.13)
$i_{775}$-dropouts per ACS field for the S1 (S2) selection criteria.
Using these distributions from GOODS, we create 10,000 Monte Carlo
(MC) quintuplets, where each MC quintuplet is generated by randomly
selecting five independent numbers of $i_{775}$-dropouts, each corresponding
to a single ACS field.  We then test how many MC quintuplets have
the counts we have observed. For the S1, we find that only $0.06 \pm 0.02$\% of the MC
quintuplets have exactly two overdense and two underdense fields.
For the S2, this probability is only $0.03 \pm 0.09$\%.
For the S3, six $i_{775}$-dropouts in one ACS field is the maximum number among the 30 ACS fields in GOODS so any MC quintuplets cannot be generated to have more than six $i_{775}$-dropouts. However, since one QSO field has 10 $i_{775}$-dropouts, we have zero probability for S3.
The error bars on these probabilities
are calculated by considering variations between 10 independent
subsets of 1,000 MC quintuplets. This comparison to GOODS empirical
dropout statistics suggests that the QSOs are indeed affecting their
environments.

Estimating the likelihood of the counts observed in our fields on
the basis of the $i_{775}$-dropout count distribution in GOODS is not
entirely appropriate as even GOODS is affected by cosmic
variance because within both the CDFS and the HDFN, the ACS fields are all
adjacent. We can use the conservative model of cosmic variance of \citet{TS08} to estimate the
likelihood of our detected counts. This model is based on extended Press-Schechter theory 
as well as synthetic catalogs extracted from N-body simulations of structure formation. 
In this  case, we establish the probability with $10^{6}$ MC quintuplets. We find that the likelihood of a MC quintuplet matching our observed distribution of over- and underdense fields
using the S1 criteria is $0.9\pm0.08\%$. S2 has a likelihood of $0.3\pm0.05\%$, 
and S3 has a likelihood of $0.8\pm0.09\%$. 
This result is less significant than that derived from the
GOODS distribution, but it is comforting that the
significance does not decrease when using samples with more
stringent color or S/N selections. Thus, while we cannot claim for
our overall sample a very significant detection of a discrepancy
from a distribution dominated by cosmic variance alone, our
distribution remains unlikely at the 99\% level.

A criticism to this type of analysis is that these are not a-priori
probabilities as we knew the outcome of the experiment before
carrying out the statistical tests. This is only partly correct
because the main idea of the HST proposal was indeed to look for
overdensities or underdensities compared to the field even though
the statistical test was not specified. Moreover, it is
possible to design an experiment that does not depend as much on the
observed counts, namely to evaluate the probability that out of the
five fields only one is within one (Poissonian) $\sigma$ of the
mean, i.e. within $8.08\pm2.84$ for selection S1, within $3.95\pm1.99$
for selection S2, or within $2.96\pm1.72$ for selection S3.
Here the formal Poisson $\sigma$ is used only to define an inner
interval and has no attached probability significance. Probabilities
are estimated by comparing how our observed object count
distribution compares to that expected from cosmic variance. We find
that the probability of finding no more than one out of five fields in the inner
interval is of 20\%  for S1, 4\% for S2, and 5.8\% for S3. The same a priori test based on the observed counts distribution in GOODS would give a probability of finding no more than one object in the inner interval of 1.5\% for S1, 0.4\% for S2, and 1.5\% for S3. 
This reinforces the view that the QSO fields have a distribution of $i_{775}$-dropout counts 
broader than what is expected by cosmic variance alone.
  
\section{CONCLUSIONS}

Summarizing our results, we find two fields where the numbers of $i_{775}$-dropout galaxies and their $i_{775}-z_{850}$ color distributions are significantly different (at 99\% confidence) than the averages for galaxies selected in the same way from GOODS fields.
When we look at the distribution of all five fields, we find that it is likely (at $80 - 96$\% confidence level, depending on selection and specific statistical test) that the
distribution of counts in the QSO fields is broader than that of
GOODS and cannot be explained by cosmic variance alone. 

We now discuss the possible implications of our results assuming
that the departure from the expected distribution of
field $i_{775}$-dropouts is indeed real. The fact that we observe both overdensities
and underdensities is somewhat puzzling. 
We know that QSOs at $z=6$ are very rare objects and are most likely associated with overdensities on large scales. Tracing a pencil beam with the area of an ACS field through a cold dark matter (CDM) simulation box with the method of Trenti \& Stiavelli (2008), we do not find correlations over $\Delta z \geq 0.3$. This is not surprising as $\Delta z=0.3$ corresponds to about 90 Mpc $h^{-1}$ at $z\sim6$ and on those scale the CDM power spectrum predicts a value of the mass fluctuation $\sigma_M$ many orders of magnitude lower than the value that can be associated with the QSO itself. From this point of view, the redshift range probed by $i_{775}$-dropouts spans at least three uncorrelated volumes.

A QSO at $z \sim 6$ is expected to live in the most massive halos within
$\approx$ Gpc$^3$ comoving volumes, with masses of the order
of $\approx 4 \times 10^{12} h^{-1} M_{\sun}$ \citep[e.g.,][]{Springel05}. Thus the dark matter halo mass function in the vicinity of
the QSO halo will be biased by the presence of a rare overdensity
\citep[e.g.,][]{Barkana04}. To quantify the impact of the QSO on
the expected number counts in its immediate neighborhood we use the
model of Munoz \& Loeb (2007). From their Fig. 4 we derive that
around the QSO there should be between 6 and 7 $i_{775}$-dropouts living in
dark halos of mass $>5 \times 10^{10} h^{-1} M_{\sun}$ taking
into account an assumed duty cycle of 0.25 for LBGs. 
The duty cycle is used to establish a halo mass scale for the observed galaxies 
by requiring that the number of halos of the required mass be equal to 
the number of objects divided by the duty cycle. Adopting a duty cycle allows us to determine a mass scale from the number of objects and to avoid using the ill-measured M/L of galaxies at z $\gta 6$. However, the results do not 
depend critically on the choice of duty cycle for range between 1 and 0.1.
Our fields do not probably reach a depth that allows us to probe these halo masses with high
completeness, but still we would expect to detect 2-3 of such LBGs
or more if the ``duty cycle'' were higher.

In this light, deficits in the number of $i_{775}$-dropout candidates are surprising. Indeed
2/3 of the expected objects are in uncorrelated volumes and
should not be affected by the presence of the QSO. The one third
affected by the QSO now becomes a very small number and detecting a
deficit in any single field is generally going to be statistically
insignificant. It is interesting to note that at the time this
project was planned the expected number of $i_{775}$-dropouts in GOODS was
thought to be higher \citep[e.g.,][]{Dickinson04} so that a deficit
would have been better quantifiable. 
Despite these considerations, the fact remains that we do seem to detect fields that have a deficit
of $i_{775}$-dropout counts compared to the field. If we really had physical
overdensities and physical underdensities near the QSO, what would
be the origin of this effect? One possible explanation is that two
physical mechanisms are simultaneously at play: the density of halos
near the QSO is indeed higher but feedback by the QSO prevents many
of these halos from becoming galaxies. The \ion{H}{2} regions generated by
luminous quasars can affect the formation and clustering of galaxies. 
\citet{Wyithe05} derived \ion{H}{2} size from displacement of quasar host galaxy redshift and the Gunn-Peterson trough redshift. The \ion{H}{2} region size of the QSO J1030+0524 is the largest of the five quasars but the second overdense field, J1630+4012 and the most underdense field,
J1306+0356, have very similar \ion{H}{2} region sizes.
The field with density comparable to GOODS,
J1048+4637, has the smallest \ion{H}{2} region size.

Thus, we find no evident correlation between density of $i_{775}$-dropouts and
\ion{H}{2} region size. This may or may not be significant as \ion{H}{2} region
sizes are roughly correlated with the luminosity of the quasars and
their lifetimes; the latter measurements are not very accurate. 
We see a weak trend between counts and QSO luminosity as the two faintest
QSOs are the two overdense ones and the most luminous QSO is one of the
underdense ones. However, the most underdense QSO field(J1306+0356) is
the third luminous QSO and within 0.04 mag from that of the most
overdense (J1030+0524). 
On the basis of these considerations we
conjecture that the suppression of galaxy formation which we may be
witnessing could be the result of percolation of ionized Hydrogen
bubbles. This would make it dependent, but not uniquely driven, by
the QSO properties. Clearly it would be desirable to study these
effects with better statistics.

Interestingly, \citet{Maselli08}, with an entirely different method, find conclusion similar to ours: namely, that J1630+4012 is overdense while J1148+5251 and J1306+0356 are underdense. They also find an overdensity around SDSS J0836+0054, also found to be overdense by \citet{Zheng06}. \citet{Maselli08} predict  that ionizing radiation from clustered galaxies for J1630+4012 exceeds the one from the quasar by a factor of five. We estimate the ionizing flux of our candidates.
The total UV flux observed by summing the $z_{850}$ photometry from all of our $i_{775}$-dropout candidates is 7.0\% and 8.5\% of the quasar flux in $z_{850}$ for J1030+0524 and J1630+4012, respectively. 
For any reasonable spectral energy distribution, the excess is too small to affect the ionizing contributions. In order to have an influence 
at this level of overdensity, the excess should span a much larger area than that provided here.
Clearly to further clarify these findings, we would need a larger sample as well as more extended data over overdense fields. 

\acknowledgments
We thank the referee for careful reading and valuable comments. This work was partially supported by HST GO grant of 01087 and 01168.
SGD and AAM acknowledge a partial support from the NSF grant AST-0407448,
and the Ajax Foundation. The work of DS was carried
out at the Jet Propulsion Laboratory, California Institute of Technology,
under a contract with NASA.

\clearpage

\begin{deluxetable}{cccccc}
\tablecaption{ Exposure times (in seconds) of QSO fields } 
\tablewidth{0pt}
\tablehead{ \colhead{Quasar Field } & {RA(J2000)} & {DEC(J2000)} &
{Redshift} & {Exptime(F775W)} & {Exptime(F850LP)} }
\startdata
J1030+0524& 10 30 27.10 & 05 24 55.0  & 6.28 & 5840 & 11330 \\
J1048+4637& 10 48 45.05 & 46 37 18.3  & 6.23 & 6130 & 11770 \\
J1148+5251& 11 48 16.64 & 52 51 50.3  & 6.40 & 6180 & 11950 \\
J1306+0356& 13 06 08.26 & 03 56 36.3  & 5.99 & 5870 & 11340 \\
J1630+4012& 16 30 33.00 & 40 12 09.6  & 6.05 & 5980 & 11580 \\
\enddata
\end{deluxetable}

\begin{deluxetable}{ccccc}
\tabletypesize{\normalsize}
\tablecaption{Number of $i_{775}$-dropouts and Poisson error by S/N ratio and color limit}
\tablewidth{0pt}
\tablehead{
\colhead{Field} & \colhead{S1\tablenotemark{a}} & \colhead{S2\tablenotemark{b}} & \colhead{S3\tablenotemark{c}} 
}
\startdata
GOODS & $8.08\pm2.84$ & $3.95\pm1.99$ & $2.96\pm1.72$  \\
J1030+0524 & $14\pm3.74$ & $8\pm2.83$ & $10\pm3.16$  \\
J1048+4637 & $8\pm2.83$ & $2\pm1.41$ & $4\pm2.00$  \\
J1148+5251 & $3\pm1.73$ & $2\pm1.41$ & $0\pm0.00$  \\
J1306+0356 & $1\pm1.00$ & $0\pm0.00$ & $1\pm1.00$  \\
J1630+4012 & $11\pm3.32$ & $8\pm2.83$ & $ 5\pm2.24$ \\
\enddata

\tablenotetext{a}{S1: S/N$>5$ and $i_{775}-z_{850}>1.3$ }
\tablenotetext{b}{S2: S/N$>5$ and $i_{775}-z_{850}>1.5$ }
\tablenotetext{c}{S3: S/N$>8$ and $i_{775}-z_{850}>1.3$ }
\tablecomments{These numbers do not include the target quasars. The
GOODS number has been normalized to the size of a single ACS field.}
\end{deluxetable}

\clearpage
\begin{deluxetable}{ccccccccc}
\tabletypesize{\scriptsize}
\tablewidth{0pt}
\tablecaption{Candidates with $i_{775}-z_{850} > 1.3$ and S/N$>5$}
\tablehead{
\colhead{Object} & \colhead{RA(J2000)} & \colhead{DEC(J2000)} & \colhead{$z_{850}$} &
\colhead{$S/N(z_{850}$)\tablenotemark{a}} & \colhead{S/N($i_{775}$)\tablenotemark{b}} &
\colhead{$i_{775}$-$z_{850}$\tablenotemark{b}} &
\colhead{$r_{hl}$(arcsec)} & \colhead{S/G\tablenotemark{c}}
}

\startdata
A1 & 10 30 21.03 & 05 24 10.13 &  26.19 &    6.19 &    3.30 &    1.30 &    0.15 &    0.02 \\
A2 & 10 30 21.57 & 05 26 07.94 &  26.33 &    6.17 &    2.96 &    1.30 &    0.15 &    0.98 \\
A3 & 10 30 24.76 & 05 24 31.65 &  25.20 &   11.94 &    5.88 &    1.30 &    0.17 &    0.03 \\
A4 & 10 30 27.79 & 05 24 31.65 &  25.66 &   10.47 &    4.91 &    1.33 &    0.16 &    0.02 \\
A5 & 10 30 27.37 & 05 23 07.30 &  25.60 &    8.81 &    4.89 &    1.36 &    0.16 &    0.03 \\
A6 & 10 30 22.66 & 05 24 37.16 &  25.72 &    8.63 &    1.85 &    1.45 &    0.31 &    0.29 \\
A7 & 10 30 22.28 & 05 24 34.51 &  26.27 &    5.64 &    1.79 &    1.52 &    0.19 &    0.19 \\
A8 & 10 30 24.98 & 05 23 34.62 &  26.38 &    9.63 &    3.59 &    1.54 &    0.08 &    0.97 \\
A9 & 10 30 19.41 & 05 24 58.25 &  26.30 &    5.17 &    2.30 &    1.60 &    0.22 &    0.25 \\
A10& 10 30 18.28 & 05 24 23.65 &  25.90 &    8.78 &    0.38 &    1.61 &    0.30 &    0.78 \\
A11& 10 30 20.62 & 05 23 43.63 &  25.56 &    8.65 &    2.69 &    1.66 &    0.26 &    0.01 \\
A12& 10 30 28.23 & 05 22 35.64 &  26.22 &    9.41 &    2.55 &    1.74 &    0.11 &    0.85 \\
A13\tablenotemark{*}& 10 30 24.08 & 05 24 20.40 &  25.74 &    9.05 &    2.59 &    2.12 &    0.14 &    0.02 \\
A14& 10 30 21.73 & 05 25 10.81 &  25.37 &    9.18 &    1.28 &    2.30 &    0.25 &    0.02 \\
QSO J1030+0524 & 10 30 27.09 & 05 24 55.00 &  20.07 &  461.80 &   45.97 &    3.16 &    0.09 &    0.85 \\
\tableline
B1&     16 30 40.71&     40 11 17.38&       26.47&        6.11&        3.12&        1.35&        0.20&        0.01\\
B2&     16 30 37.55&     40 11 55.94&       25.98&       13.55&        7.07&        1.44&        0.09&        0.98\\
B3&     16 30 40.55&     40 12 20.01&       25.50&        9.90&        4.35&        1.47&       0.34&        0.02\\
B4&     16 30 40.51&     40 12 43.29&       25.01&       13.49&        4.57&        1.62&       0.47&        0.02\\
B5&     16 30 26.62&     40 13 31.02&       25.92&        7.68&        3.12&        1.71&        0.27&        0.01\\
B6&     16 30 27.90&     40 11 16.90&       26.31&        7.73&        3.46&        1.79&        0.15&        0.02\\
B7&     16 30 43.82&     40 11 59.92&       26.16&        7.88&        1.41&        1.86&        0.25&        0.01\\
B8&     16 30 26.87&     40 12 45.88&       26.21&        7.31&        0.45&        1.87&        0.25&        0.02\\
B9&     16 30 41.11&     40 13 5.89&       26.16&        7.84&        0.80&        2.12&        0.24&        0.01\\
B10&     16 30 42.48&     40 12 14.05&       25.93&        9.46&        2.47&        2.24&        0.18&        0.02\\
B11&     16 30 36.99&     40 13 9.27&       26.30&       10.28&        1.51&        2.26&        0.17&        0.02\\
QSO J1630+4012&     16 30 33.85&     40 12 9.48&       20.63&      338.76&       83.90&        0.10&        0.10&        0.87\\
\tableline
C1 & 10 48 52.34 & 46 36 12.29 &  26.46 &    5.62 &    2.30 &    1.30 &    0.17 &    0.83 \\
C2 & 10 48 42.39 & 46 36 40.56 &  26.14 &    6.36 &    1.77 &    1.34 &    0.21 &    0.78 \\
C3 & 10 48 52.48 & 46 37 11.05 &  24.38 &   12.80 &    7.88 &    1.36 &    0.36 &    0.02 \\
C4 & 10 48 53.47 & 46 35 55.93 &  25.19 &    8.45 &    4.60 &    1.36 &    0.24 &    0.02 \\
C5 & 10 48 42.21 & 46 38 20.34 &  25.09 &   10.65 &    3.02 &    1.37 &    0.37 &    0.17 \\
C6 & 10 48 47.62 & 46 36 01.41 &  26.21 &    5.60 &    2.10 &    1.64 &    0.18 &    0.11 \\
C7 & 10 48 42.13 & 46 38 21.58 &  24.53 &   12.42 &    1.98 &    1.82  &    0.55 &    0.01 \\
QSO J1048+4637& 10 48 45.22 & 46 37 17.92 & 19.98 &  303.81 &   61.30 &    2.92 &    0.11 &    0.86 \\
\tableline
D1 & 11 48 07.93 & 52 51 59.32 &  26.07 &    7.03 &    3.52 &    1.40 &    0.13 &    0.84 \\
D2 & 11 48 15.41 & 52 51 07.14 &  26.08 &    6.72 &    1.88 &    1.52 &    0.25 &    0.49 \\
D3 & 11 48 20.29 & 52 53 04.22 &  26.28 &    7.61 &    1.59 &    2.07 &    0.14 &    0.01 \\
QSO J1148+5251 & 11 48 16.74 & 52 51 50.11 & 19.83 &  450.77 &   66.60 &    3.03 &    0.10 &    0.85 \\
\tableline
E1 & 13 06  2.08 & 03 56 04.70  &  25.36 &  8.69 &   -0.19 &    1.44 &    0.30 &    0.13 \\
QSO J1306+0356& 13 06 08.22 & 03 56 25.92 &  20.00 &  272.52 &   97.22 &    2.23 &    0.10 &    0.85 \\

\enddata
\tablenotetext{a}{Signal to noise ratio calculated using FLUX\_AUTO.}
\tablenotetext{b}{Calculated with FLUX\_ISO}
\tablenotetext{c}{Star-Galaxy classification index from SExtractor}
\tablenotetext{*}{Spectroscopically confirmed; $z=5.97$ \citet{Stiavelli05}}
\end{deluxetable}

\clearpage

\begin{figure}
\epsscale{.80}
\plotone{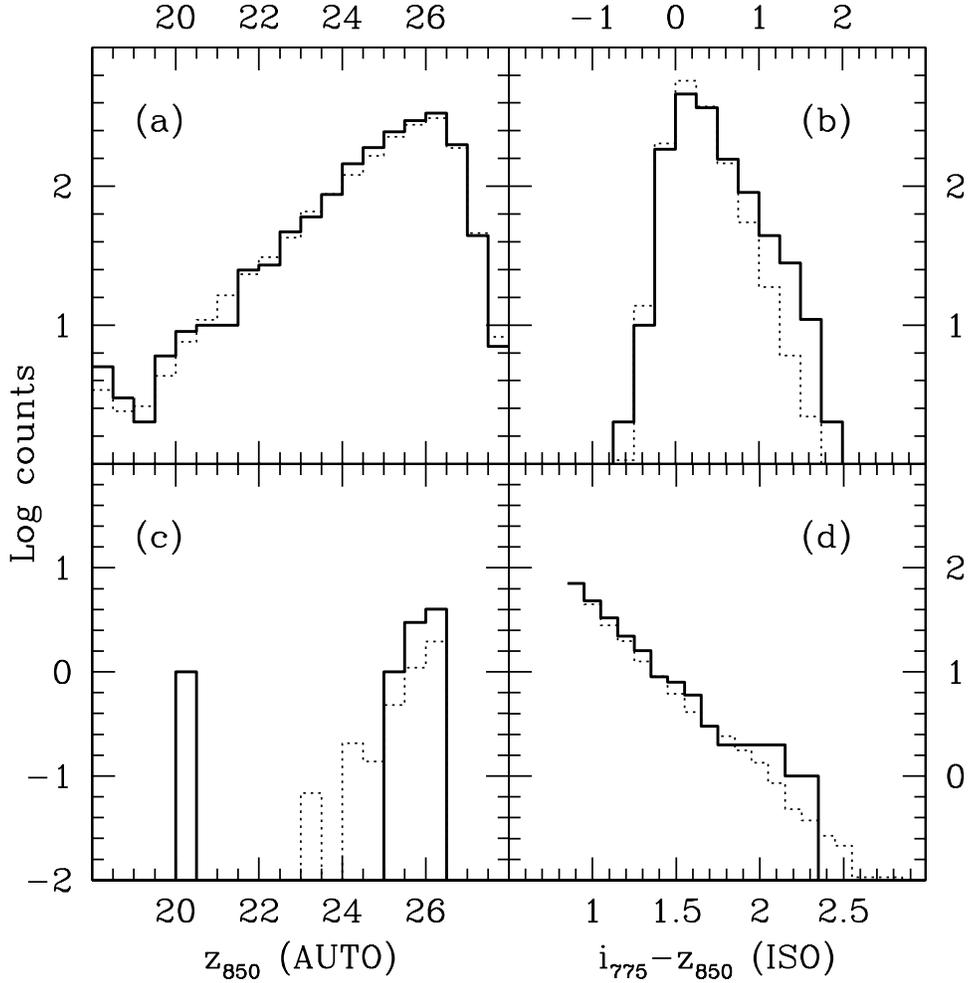} 
\caption{ Distributions of the number of objects
versus magnitude and color, comparing objects in the SDSS J1030+5032
field (solid histograms) to those in the GOODS field (dotted
histograms). The GOODS counts are normalized to the area of the
quasar field ($\sim$11.3 arcmin$^{2}$). Panel (a) shows the total
counts vs. $z_{850}(AUTO)$ of all objects with no
selection criteria applied. Panel (b) shows the total counts vs.
$i_{775}$-$z_{850}$ of all objects with no selection
criteria applied. Panel (c) shows the candidates with
$i_{775}$-$z_{850}>1.5$; the brightest object is the target QSO.
Panel (d) shows the number of objects redder than a given
$i_{775}$-$z_{850}$, excluding the target QSO, with the GOODS counts renormalized to the QSO counts at $0.9<i_{775}$-$z_{850}<1.0$.}
\end{figure}
\clearpage

\begin{figure}
\epsscale{.80}
\plotone{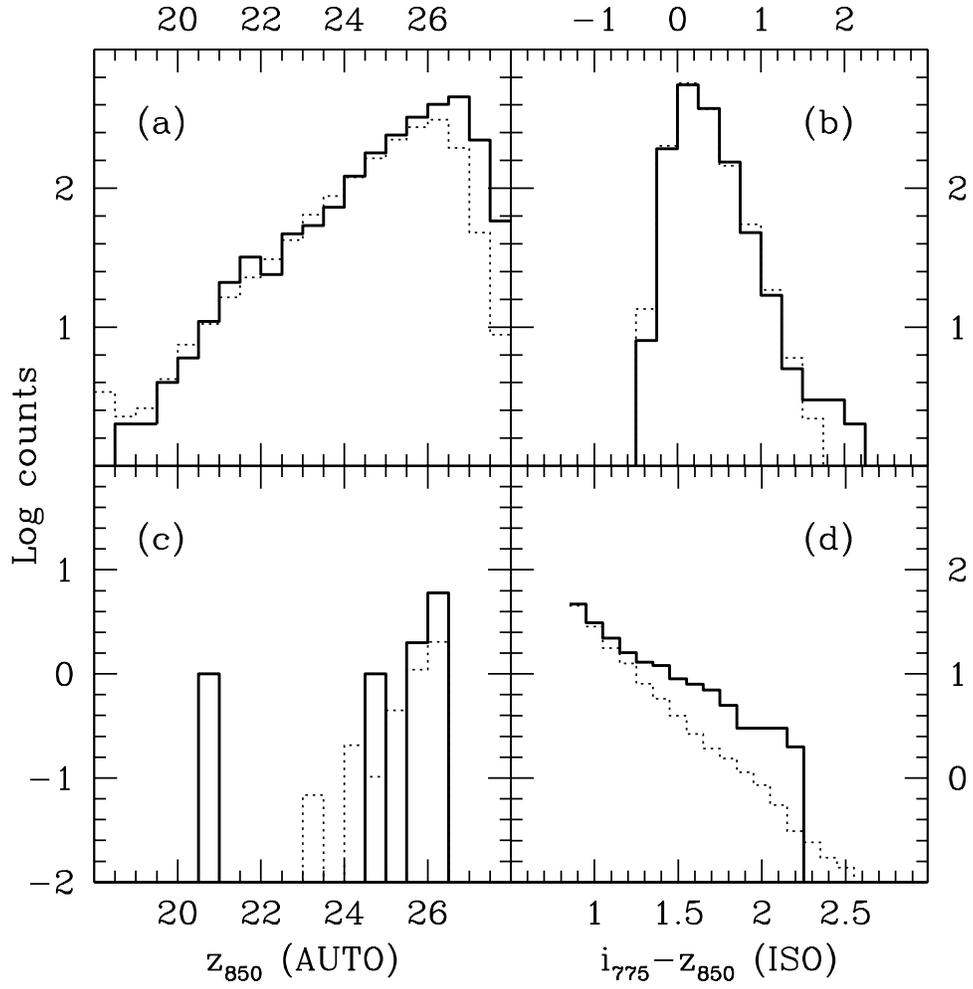}
\caption{ Same as Fig 1. for J1630+4012 }
\end{figure}
\clearpage

\begin{figure}
\epsscale{.80}
\plotone{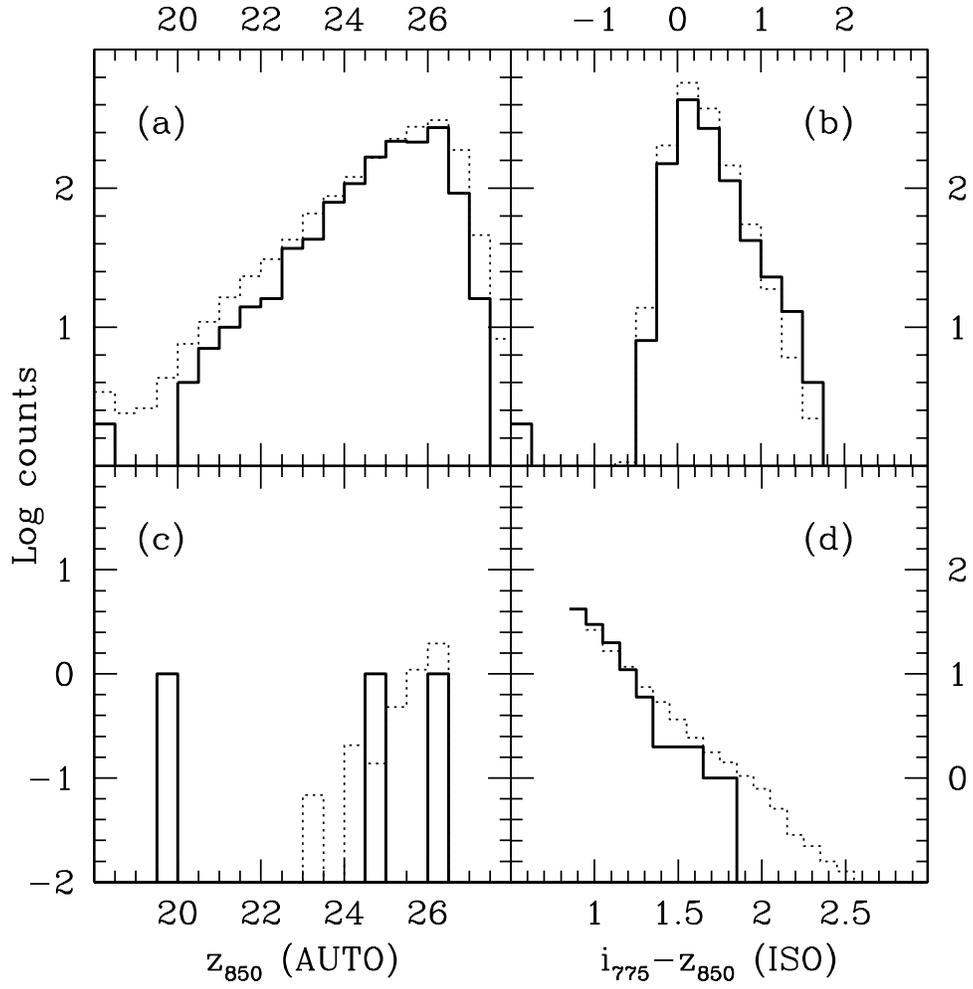}
\caption{ Same as Fig 1. for J1048+4637 }
\end{figure}
\clearpage

\begin{figure}
\epsscale{.80}
\plotone{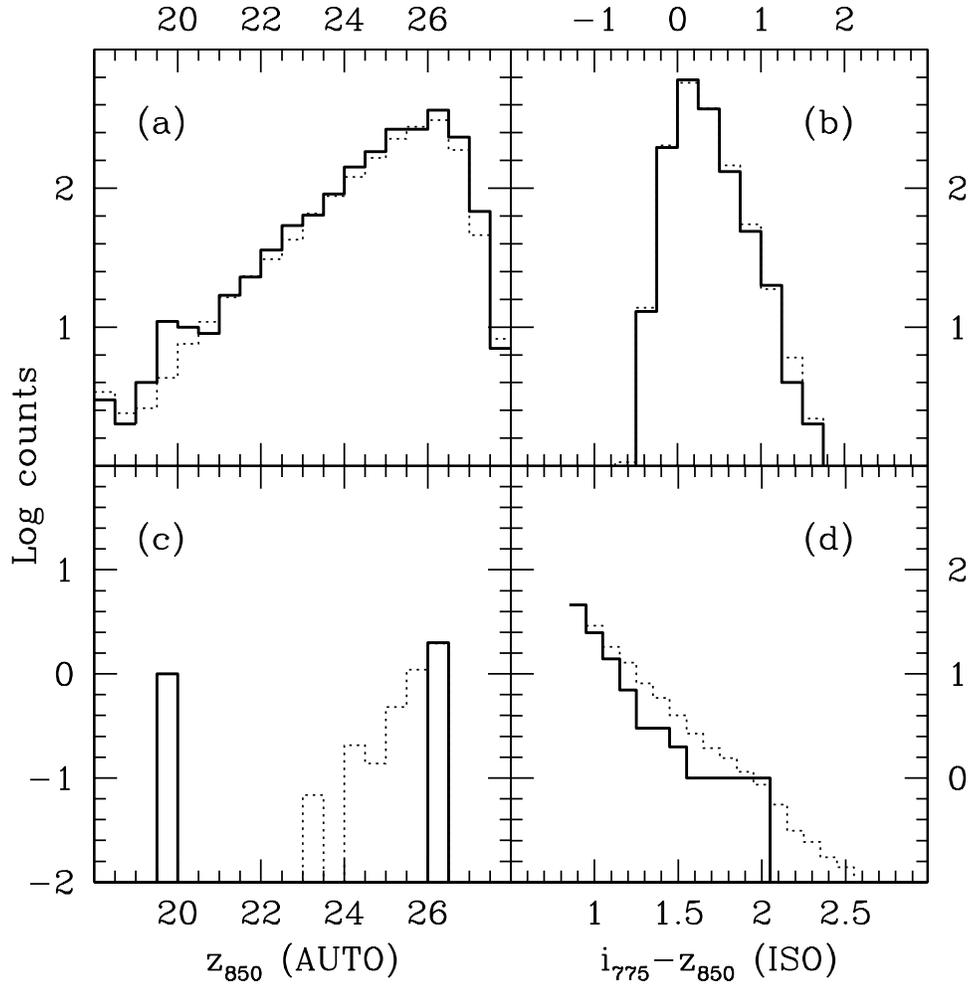}
\caption{ Same as Fig 1. for J1148+5251  }
\end{figure}
\clearpage

\begin{figure}
\epsscale{.80}
\plotone{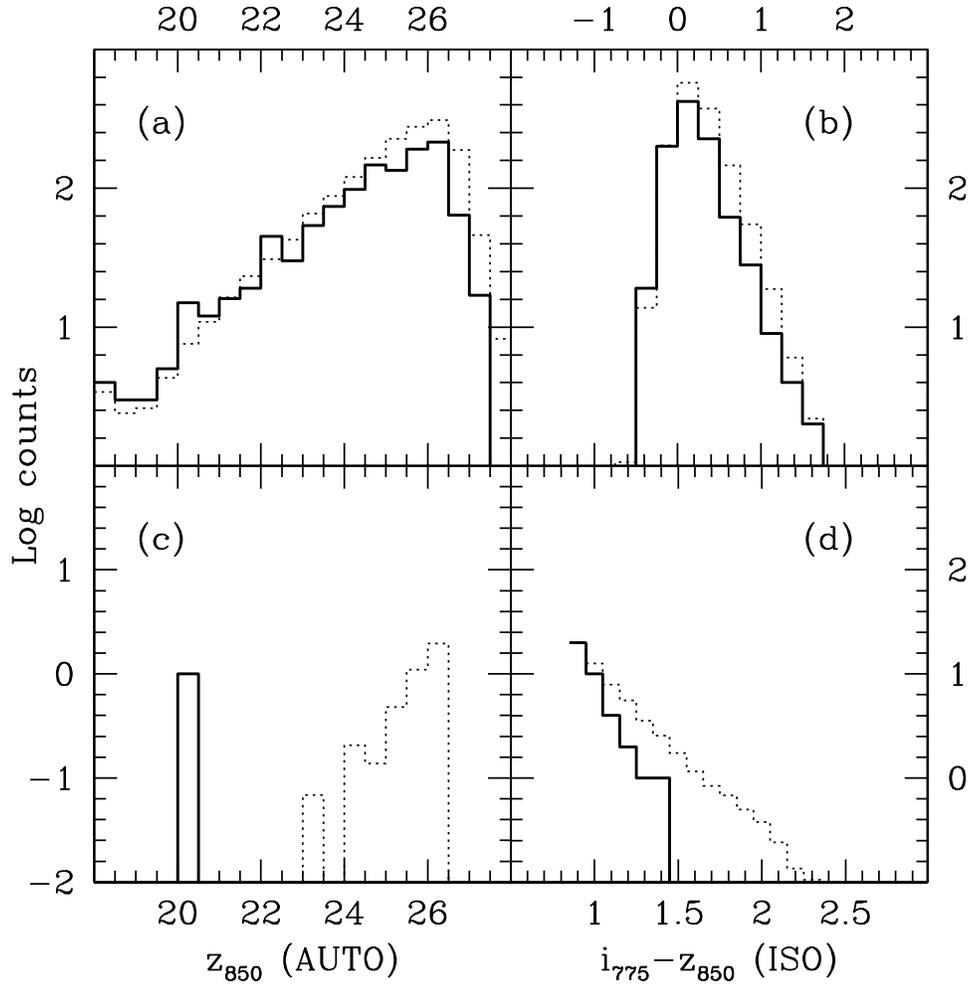}
\caption{ Same as Fig 1. for J1306+0356 }
\end{figure}
\clearpage

\begin{figure}
\epsscale{0.8}
\plotone{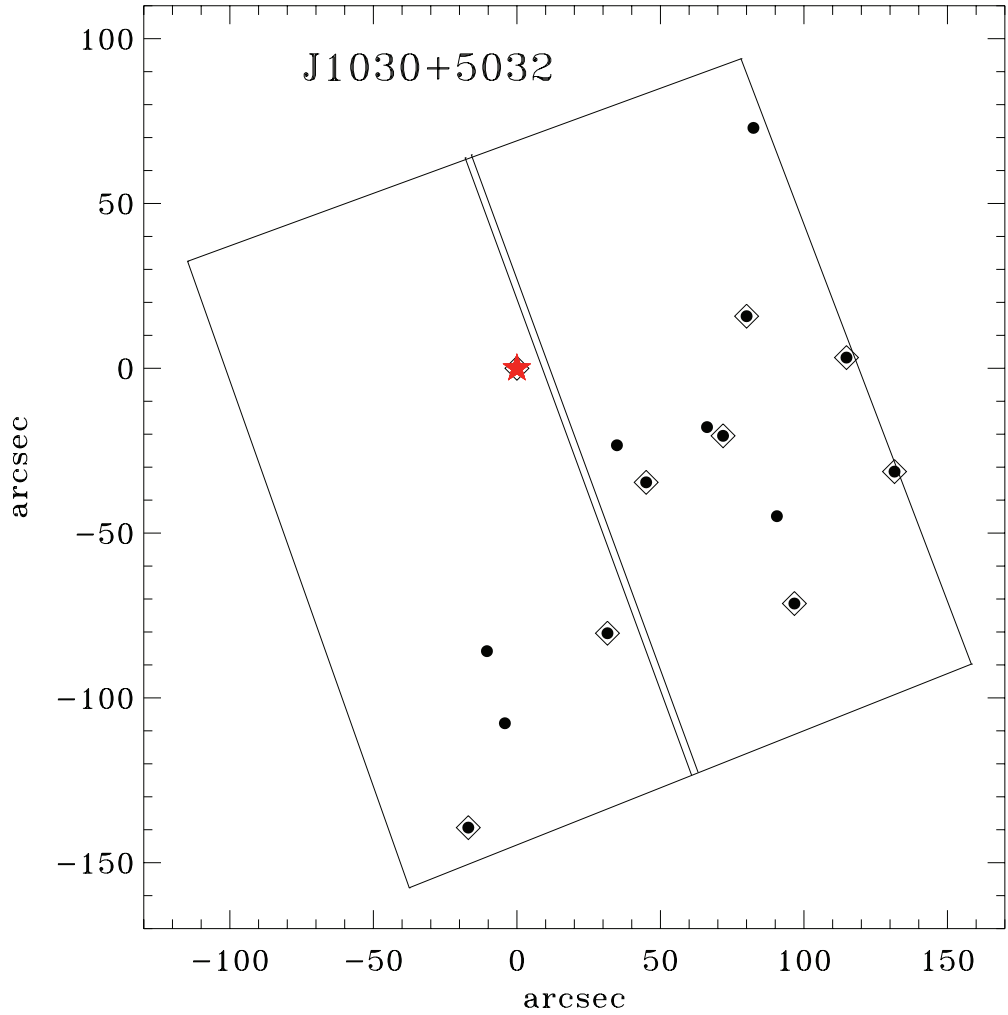}
\caption{ The spatial distribution of $i_{775}$-dropouts redder than 1.3 (circles) and
redder than 1.5 (open squares) for S/N$>5$ in the J1030+0524 field.
The star represents the QSO SDSS J1030+0524. The axes of x and y are in arcseconds and they are relative to the QSO position. East is to the left and north is up.}
\end{figure}
\clearpage

\begin{figure}
\epsscale{0.8}
\plotone{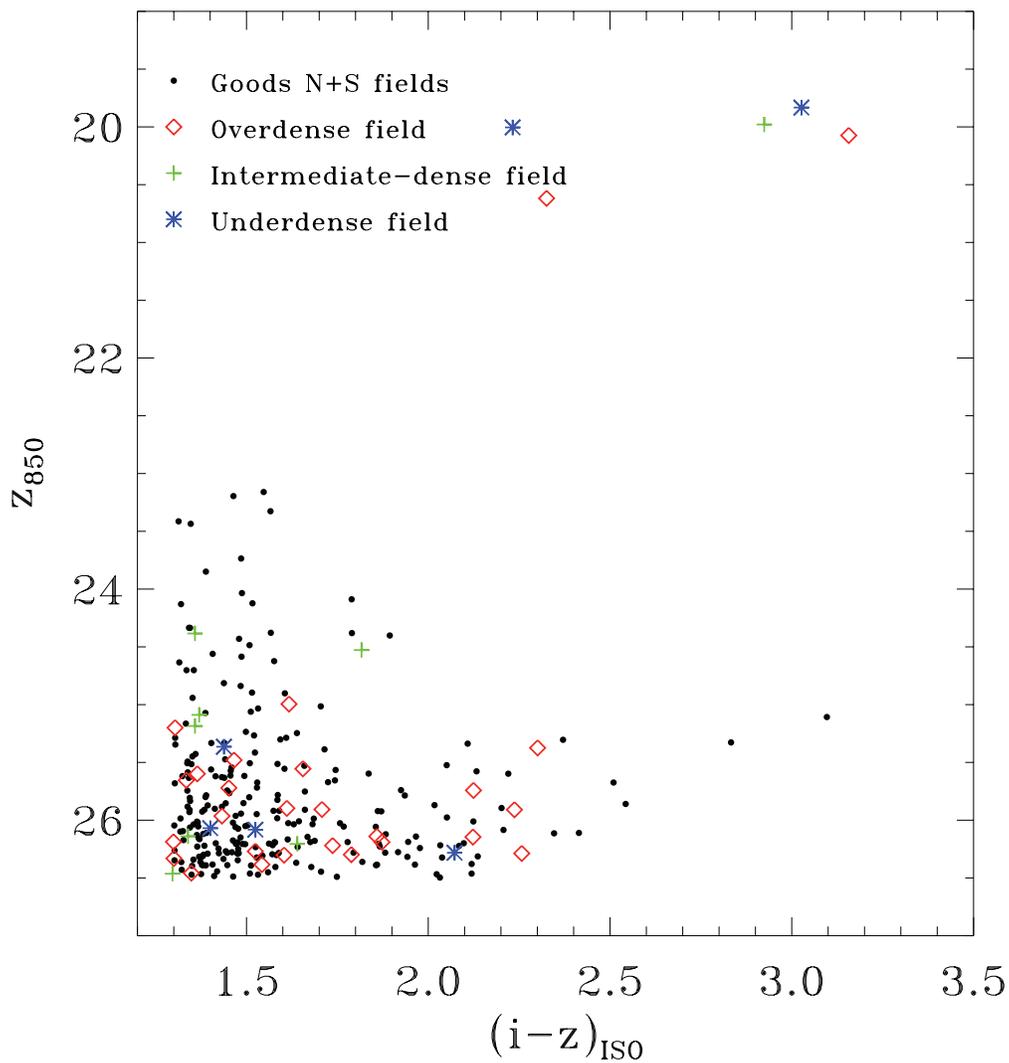}
\caption{ Color-magnitude distribution of $i_{775}$-dropouts, selected using S1 criteria, in GOODS (small black dots) and the five QSO fields. The overdense QSO fields are indicated with red diamonds, the underdense QSO fields are indicated with blue asterisks, and the intermediate density QSO field is indicated with green crosses.
The brightest objects, with $z_{850} \lta 21$, are the QSOs.}
\end{figure}
\clearpage

\begin{figure}
\epsscale{0.8}
\plotone{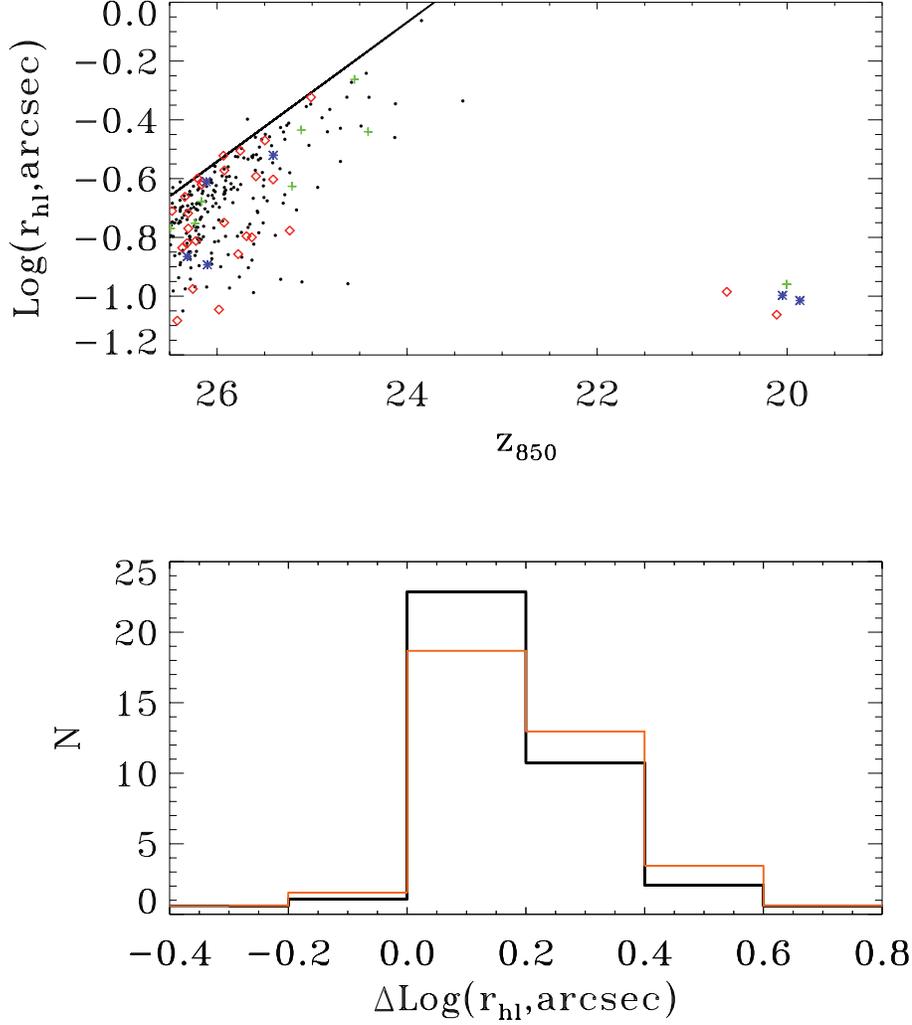} 
\caption{ The {\it upper panel} shows half-light radii
of $i_{775}$-dropout candidates (S1 sample) with respect to $z_{850}$ magnitude for excess (red
diamonds), deficit (blue asterisks), and intermediate (green crosses) density fields, as well as for GOODS (black dots). The five bright objects at $z_{850} \lta 21$ are the QSOs. The solid line is fitted to the upper limit of the logarithmic radii of the objects in GOODS. The {\it lower
panel} shows the histogram of the distances from the
upper envelope on the upper panel to the logarithmic radii data of the
all five QSO fields (orange line) and the GOODS field (black line) along the axis of ordinates. The histogram of GOODS is normalized to the area of the QSO fields.}
\end{figure}
\clearpage

\begin{figure}
\epsscale{0.8}
\plotone{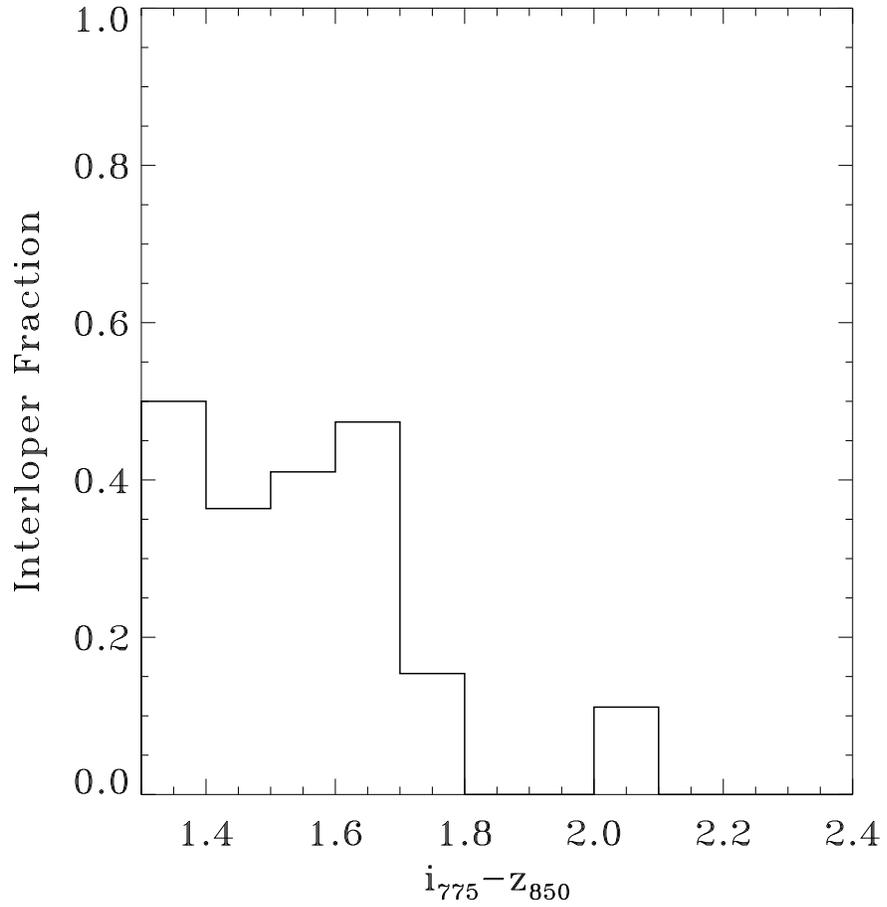}
\caption{The ratio of the number of GOODS $i_{775}$-dropouts selected by us (S1) but rejected by the full GOODS selection criteria, including S/N$(V_{606})<2.0$ or $V_{606}-z_{850}<2.8$  \citep{Beckwith06} to the number of GOODS $i_{775}$-dropouts selected by our criteria.  }
\end{figure}
\clearpage

\begin{figure}
\epsscale{0.8}
\plotone{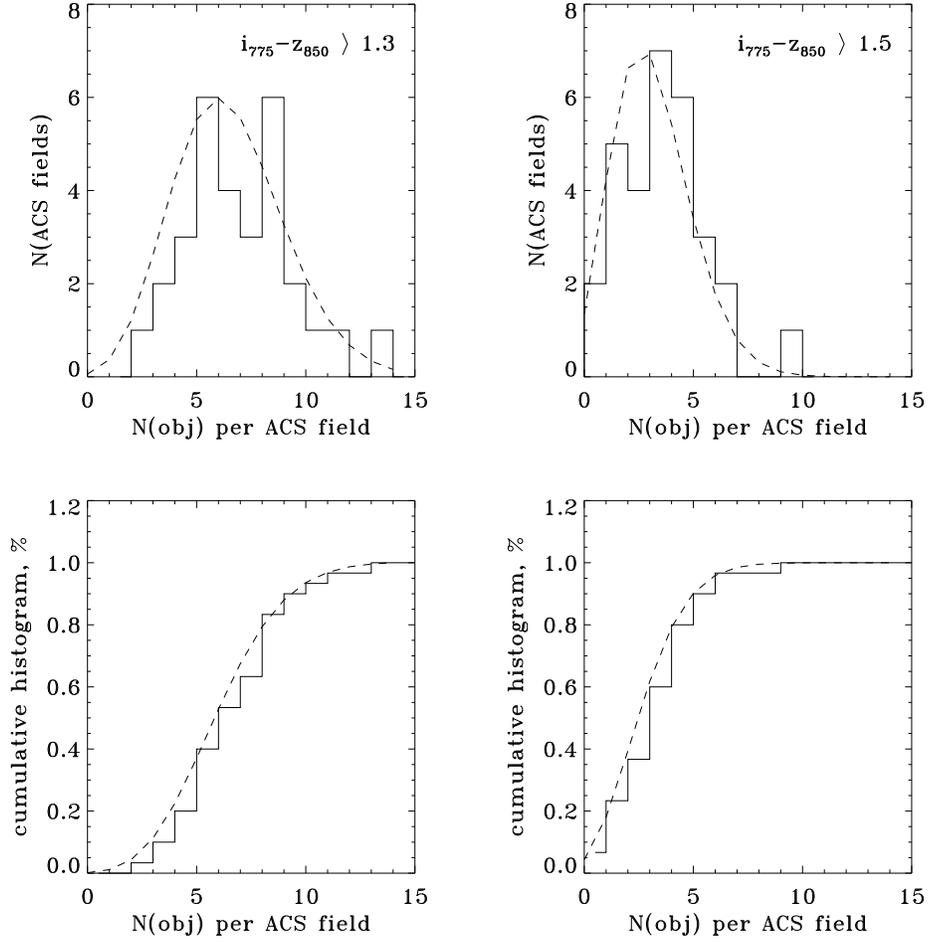}
\caption{ Histograms ({\it upper panels})
of the number of $i_{775}$-dropouts per ACS field in the GOODS North and
South (Solid lines) for S1 sample (on the left) and S2 sample (on the
right). Poisson distributions (dashed lines) with a mean of 6.5 for $i_{775}-z_{850}>1.3$ and 3.13 for $i_{775}-z_{850}> 1.5$ are fitted. The 30 ACS fields are overlaid in the whole GOODS field. The {\it bottom panels} show cumulative distributions from the histograms}
\end{figure}
\clearpage


\begin{thebibliography}{}

\bibitem[Barkana \& Loeb(2004)]{Barkana04} Barkana, R., \& Loeb,
A.\ 2004, \apj, 609, 474

\bibitem[Barth et al.(2003)]{Barth03} Barth, A.~J., Martini, 
P., Nelson, C.~H., \& Ho, L.~C.\ 2003, \apjl, 594, L95 

\bibitem[Becker et al.(2001)]{Becker01} Becker, R.~H., et al.\
2001, \aj, 122, 2850


\bibitem[Beckwith et al.(2006)]{Beckwith06} Beckwith, S.~V.~W., et
al.\ 2006, \aj, 132, 1729


\bibitem[Bertin \& Arnouts(1996)]{Bertin96} Bertin, E., \&
Arnouts, S.\ 1996, \aaps, 117, 393

\bibitem[Bolton \& Haehnelt(2007)]{Bolton07} Bolton, J.~S., \& Haehnelt, M.~G.\ 2007, \mnras, 374, 493


\bibitem[Bouwens et al.(2007)]{Bouwens07} Bouwens, R.~J.,
Illingworth, G.~D., Franx, M., \& Ford, H.\ 2007, \apj, 670, 928

\bibitem[Bunker et al.(2004)]{Bunker04} Bunker, A.~J., Stanway,
E.~R., Ellis, R.~S., \& McMahon, R.~G.\ 2004, \mnras, 355, 374

\bibitem[Cen(2003)]{Cen03} Cen, R.\ 2003, \apj, 591, 12

\bibitem[Ciardi et al.(2003)]{Ciardi03} Ciardi, B., Ferrara, A.,
\& White, S.~D.~M.\ 2003, \mnras, 344, L7

\bibitem[Dickinson et al.(2004)]{Dickinson04} Dickinson, M., et
al.\ 2004, \apjl, 600, L99


\bibitem[Djorgovski(1999)]{Djorgo99} Djorgovski, S.~G.\ 1999,
The Hy-Redshift Universe: Galaxy Formation and Evolution at High Redshift,
193, 397


\bibitem[Djorgovski et al.(2001)]{Djorgovski01} Djorgovski, S.~G.,
Castro, S., Stern, D., \& Mahabal, A.~A.\ 2001, \apjl, 560, L5


\bibitem[Djorgovski et al.(1999)]{Djorgovski99} Djorgovski, S.~G.,
Odewahn, S.~C., Gal, R.~R., Brunner, R.~J., \& de Carvalho, R.~R.\ 1999,
Photometric Redshifts and the Detection of High Redshift Galaxies, 191, 179


\bibitem[Djorgovski et al.(2003)]{Djorgovski03} Djorgovski, S.~G.,
Stern, D., Mahabal, A.~A., \& Brunner, R.\ 2003, \apj, 596, 67


\bibitem[Efstathiou \& Rees(1988)]{Efstathiou88} Efstathiou, G., \&
Rees, M.~J.\ 1988, \mnras, 230, 5P


\bibitem[Fan et al.(2001)]{Fan01} Fan, X., et al.\ 2001, \aj,
122, 2833

\bibitem[Fan et al.(2003)]{Fan03} Fan, X., et al.\ 2003, \aj, 
125, 1649 

\bibitem[Fan et al.(2006)]{Fan06} Fan, X., et al.\ 2006, \aj,
132, 117


\bibitem[Fruchter \& Hook(2002)]{Fruchter02} Fruchter, A.~S., \&
Hook, R.~N.\ 2002, \pasp, 114, 144


\bibitem[Giavalisco et al.(2004)]{Giavalisco04} Giavalisco, M., et
al.\ 2004, \apjl, 600, L93

\bibitem[Gnedin(2004)]{Gnedin04} Gnedin, N.~Y.\ 2004, \apj, 610,
9

\bibitem[Gnedin \& Ostriker(1997)]{Gnedin97} Gnedin, N.~Y., \&
Ostriker, J.~P.\ 1997, \apj, 486, 581


\bibitem[Gunn \& Peterson(1965)]{Gunn65} Gunn, J.~E., \&
Peterson, B.~A.\ 1965, \apj, 142, 1633


\bibitem[Haiman \& Holder(2003)]{Haiman03} Haiman, Z., \&
Holder, G.~P.\ 2003, \apj, 595, 1

\bibitem[Kaiser(1984)]{Kaiser84} Kaiser, N.\ 1984, \apjl, 284, L9

\bibitem[Koekemoer et al.(2002)]{Koekemoer02} Koekemoer, A.~M.,
Fruchter, A.~S., Hook, R.~N., \& Hack, W.\ 2002, The 2002 HST Calibration
Workshop : Hubble after the Installation of the ACS and the NICMOS Cooling
System, Proceedings of a Workshop held at the Space Telescope Science
Institute, Baltimore, Maryland, October 17 and 18, 2002.~ Edited by
Santiago Arribas, Anton Koekemoer, and Brad Whitmore.~Baltimore, MD: Space
Telescope Science Institute, 2002., p.337, 337

\bibitem[Komatsu et al.(2008)]{Komatsu08} Komatsu, E., et al.\
2008, ArXiv e-prints, 803, arXiv:0803.0547

\bibitem[Lidz et al.(2006)]{Lidz06} Lidz, A., Oh, S.~P.,
\& Furlanetto, S.~R.\ 2006, \apjl, 639, L47

\bibitem[Malhotra et al.(2005)]{Malhotra05} Malhotra, S., et al.\
2005, \apj, 626, 666

\bibitem[Maselli et al.(2008)]{Maselli08} Maselli, A., et al.\
2008, \mnras, submitted

\bibitem[McDonald \& Miralda-Escud{\'e}(2001)]{McDonald01}
McDonald, P., \& Miralda-Escud{\'e}, J.\ 2001, \apjl, 549, L11


\bibitem[Miralda-Escud{\'e} et al.(2000)]{Miralda00}
Miralda-Escud{\'e}, J., Haehnelt, M., \& Rees, M.~J.\ 2000, \apj, 530, 1


\bibitem[Mu{\~n}oz \& Loeb(2008)]{Mu07} Mu{\~n}oz, J.~A., \& Loeb, A.\ 2008, \mnras, 385, 2175

\bibitem[Oesch et al.(2007)]{Oesch07} Oesch, P.~A., et al.\
2007, \apj, 671, 1212

\bibitem[Ouchi et al.(2005)]{Ouchi05} Ouchi, M., et al.\ 2005, 
\apjl, 620, L1 

\bibitem[Schlegel et al.(1998)]{Schlegel98} Schlegel, D.~J.,
Finkbeiner, D.~P., \& Davis, M.\ 1998, \apj, 500, 525

\bibitem[Shapiro \& Raga(2001)]{Shapiro01} Shapiro, P.~R., \&
Raga, A.~C.\ 2001, Revista Mexicana de Astronomia y Astrofisica Conference
Series, 10, 109

\bibitem[Shull \& Venkatesan(2007)]{Shull07} Shull, M., \&
Venkatesan, A.\ 2007, ArXiv Astrophysics e-prints, arXiv:astro-ph/0702323

\bibitem[Somerville et al.(2003)]{Somerville03} Somerville, R.~S.,
Bullock, J.~S., \& Livio, M.\ 2003, \apj, 593, 616

\bibitem[Springel et al.(2005)]{Springel05} Springel, V., et al.\ 
2005, \nat, 435, 629

\bibitem[Steidel et al.(2003)]{Steidel03} Steidel, C.~C.,
Adelberger, K.~L., Shapley, A.~E., Pettini, M., Dickinson, M., \&
Giavalisco, M.\ 2003, \apj, 592, 728

\bibitem[Stiavelli et al.(2005)]{Stiavelli05} Stiavelli, M., et
al.\ 2005, \apjl, 622, L1

\bibitem[Stiavelli et al.(2004)]{Stiavelli04} Stiavelli, M., Fall,
S.~M., \& Panagia, N.\ 2004, \apjl, 610, L1

\bibitem[Trenti \& Stiavelli(2007)]{TS07} Trenti, M., \&
Stiavelli, M.\ 2007, \apj, 667, 38

\bibitem[Trenti \& Stiavelli(2008)]{TS08} Trenti, M., \&
Stiavelli, M.\ 2008, \apj, 676, 767

\bibitem[Venemans et al.(2003)]{Venemans03} Venemans, B.~P., Kurk,
J.~D., Miley, G.~K., \& R{\"o}ttgering, H.~J.~A.\ 2003, New Astronomy
Review, 47, 353


\bibitem[White et al.(2003)]{White03} White, R.~L., Becker,
R.~H., Fan, X., \& Strauss, M.~A.\ 2003, \aj, 126, 1

\bibitem[Willott et al.(2005)]{Willott05} Willott, C.J., et al.,
  2005, in ``Growing Black Holes'', ESO (Garching), Merloni et
  al. eds,  (also astro-ph/0410306)

\bibitem[Wyithe \& Loeb(2003)]{Wyithe03} Wyithe, J.~S.~B., \&
Loeb, A.\ 2003, \apjl, 588, L69


\bibitem[Wyithe et al.(2005)]{Wyithe05} Wyithe, J.~S.~B., Loeb,
A., \& Carilli, C.\ 2005, \apj, 628, 575


\bibitem[Yan \& Windhorst(2004)]{Yan04} Yan, H., \&
Windhorst, R.~A.\ 2004, \apjl, 600, L1

\bibitem[Zheng et al.(2006)]{Zheng06} Zheng, W., et al.\ 2006,
\apj, 640, 574

\bibitem[Zombeck(1990)]{Zombeck90} Zombeck, M.V. 1990, Handbook of Space Astronomy and Astrophysics
(Cambridge University Press: Cambridge), p.77


\end{thebibliography}
\end{document}